\documentclass[11pt]{book}
\usepackage{amsmath}
\usepackage{amssymb}
\usepackage{epsfig}
\usepackage{url}
\renewcommand{\thefootnote}{\ensuremath{\fnsymbol{footnote}}}

\font\rs=cmss10.360pk
\font\rt=cmss9.360pk
\font\sd=cmcsc9.360pk

\begin{document}
\oddsidemargin 0truemm
\evensidemargin 0truemm
\topmargin -10truemm

\thispagestyle{plain}

\noindent{\rs Applied Mathematical and Computational Sciences}

\vspace{-0.25cc}

\noindent{\scriptsize Vol. 1, No. 2 (2010), pp.~181--197.}

\vspace{5cc}

\begin{center}
{\Large\bf FIRST- AND SECOND-ORDER\\ WAVE GENERATION THEORY
\rule{0mm}{6mm}\renewcommand{\thefootnote}{}\footnotetext{\hspace{-0.55cm}\scriptsize
{\rm Received 4 May 2010, Accepted 30 June 2010.}
\\
{\it 2010 Mathematics Subject Classification}: 76B15, 74J15, 74J30, 34B05, 34B15, 34K10, 35G30.
\\
{\it Key words and Phrases}: boundary value problem
(\textsc{bvp}), surface waves, signalling problem, first-order and
second-order steering.}}

\vspace{2cc}

{\large\sc Natanael Karjanto}

\vspace{2cc}

\parbox{34cc}{{\small{\bf Abstract.}
The first-order and the second-order wave generation theory is
studied in this paper. The theory is based on the fully nonlinear
water wave equations. The nonlinear boundary value problem
(\textsc{bvp}) is solved using a series expansion method. Using
this method, the problem becomes a set of linear, signalling
problems according to the expansion order. The first-order theory
leads to a homogeneous \textsc{bvp}. 
It is a system with the first-order steering of the wavemaker motion as input 
and the surface wave field with propagating and evanescent modes as output. 
The second-order theory leads to a nonhomogeneous \textsc{bvp}. It is
a system where the second-order steering of the wavemaker motion
is prescribed in such a way that the second-order part of the
surface elevation far from the wavemaker contains only the bound
wave component and the free wave component vanishes. The
second-order surface wave elevation consists of a superposition of
bichromatic frequencies. \par}}
\end{center}

\vspace{1.5cc}

\begin{center}
{\bf 1. INTRODUCTION}
\end{center}

In this paper, we will consider the problem how to generate waves
in a wave tank of a hydrodynamic laboratory. The wave tank in the
context of this paper is a facility with a wavemaker on one side
and an artificial wave absorbing beach on the other side. We consider a tank 
with a flat bottom, and no water is flowing in or out of the tank.
Typically, the situation is that waves are generated by a flap
type wavemaker at one side of a (long) tank; the motion of the
flap `pushes' the waves to start propagating along the tank. This
means that typically, we are dealing with a signalling problem or a
boundary value problem (\textsc{bvp}), which is different from an
initial value problem (\textsc{ivp}) when one tries to find the
evolution of waves from given surface elevation and velocities at
an initial moment. To illustrate this for the simplest possible
case, consider the linear, non-dispersive second-order wave
equation for waves in one spatial direction $x$ and time $t:$
$\partial_{t}^{2}\eta = c^{2}\partial_{x}^{2}\eta$. Here, $\eta$
denotes the surface wave elevation, and $c > 0$ is the constant
propagation speed. The general solution is given by $\eta(x,t) =
f(x - ct) + g(x + ct)$, for arbitrary functions $f$ and $g$. The
term $f(x - ct)$ is the contribution of waves travelling to the
right (in the positive $x$-direction), and $g(x + ct)$ waves running
to the left. For the \textsc{ivp}, specifying at an initial time
(say at $t=0$) the wave elevation $\eta(x,0)$ and the velocity
$\partial_{t}\eta(x,0)$ determines the functions $f$ and $g$
uniquely. For the \textsc{bvp}, resembling the generation at
$x=0$, we prescribe the wave elevation at $x=0$ for all positive
time, assuming the initial elevation to be zero for positive $x$
(in the tank, this means a flat surface prior to the start of the
generation). If the signal is given by $s(t)$, vanishing for $t <
0$, the corresponding solution running into the tank is $\eta(x,t)
= f(x - ct)$ which should be equal to $s(t)$ at $x = 0$, leading
to $\eta (x,t) = s(t - x/c)$. Reversely, for a desired wave field
$f(x - ct)$ running in the tank, the required surface elevation at
$x = 0$ is given by $s(t) = f(-ct)$. This shows the characteristic
property of the \textsc{bvp} for the signalling problem.

The actual equations for the water motion and the precise
incorporation of the flap motion are much more difficult than
shown in the simple example above. In particular, for the free
motion of waves, there are two nontrivial effects. The first effect
is dispersion: the propagation speed of waves depends on their
wavelength (or frequency), described explicitly in the linear
theory (i.e. for small surface elevations) by the linear
dispersion relation (\textsc{ldr}), see formula (\ref{ldr}). In
fact, for a given frequency, there is one normal mode that travels
to the right as a harmonic wave, the propagating mode, and there
are solutions decaying exponentially for increasing distance (the
evanescent modes). We will use the right propagating mode and the
evanescent modes as building blocks to describe the generation of
waves by the wavemaker. For each frequency in the spectrum, the
Fourier amplitude of the flap motion is then related to the
amplitude of the corresponding propagating and evanescent modes.
If we assume these amplitudes to be sufficiently small, say of the
`first order' $\epsilon$, with $\epsilon$ is a small quantity, we
will be able to deal with nonlinear effects in a sequential way.
This is needed because the second effect is that in reality, the
equations are nonlinear. The quadratic nature of the nonlinearity
implies that each two wave components will generate other
components with an amplitude that is proportional to the product
of the two amplitudes, the so-called `bound wave components' which
have amplitudes of the order $\epsilon^{2}$. These are the
so-called `second-order effects'. For instance, two harmonic waves
of frequency $\omega_{1},\omega_{2}$ and wavenumber $k_{1}, k_{2}$
(related by the \textsc{ldr}) will have a bound wave with
frequency $\omega_{1} + \omega_{2}$ and wavenumber $k_{1} +
k_{2}$. Since the \textsc{ldr} is a concave function of the
wavenumber, this last frequency-wavenumber combination does not
satisfy the \textsc{ldr}, i.e. this is not a free wave: it can
only exist in the combination of the free wave. This
second-order bound wave that comes with a first-order free wave
has also its consequence for the wave generation. If the
first-order free-wave component is compatible with the flap
motion, the presence of the bound wave component will disturb the
wave motion, such that the additional second-order free-wave will
be generated as well. This is undesired, since the second-order
free wave component has a different propagation speed as the bound
wave component, thereby introducing a spatially inhomogeneous wave
field. That is why we add to the flap motion the additional
effects of second-order bound waves, thereby preventing any
second-order free-wave component to be generated. This process is
called `\textsl{second-order steering}' of the wavemaker motion.

This technique can be illustrated using a simple \textsc{ivp} for
an ordinary differential equation as follows. Consider the
nonlinear equation with a linear operator $\cal{L}$
\begin{equation*}
\textmd{$\cal{L}$} \eta := \partial_{t}^{2} \eta + \omega_{0}^{2}
\eta = \eta^{2},
\end{equation*}
for which we look for small solutions, say of order $\epsilon$, a
small quantity. The series expansion technique then looks for a
solution in the form
\begin{equation*}
\eta = \epsilon \eta^{(1)} + \epsilon^{2} \eta^{(2)} +
\textmd{$\cal{O}$}(\epsilon^{3}).
\end{equation*}
Substitution in the equation and requiring each order of
$\epsilon$ to vanish leads to a sequence of \textsc{ivp}s, the
first two of which read:
\begin{equation*}
\textmd{$\cal{L}$} \eta^{(1)} = 0; \qquad \textmd{$\cal{L}$}
\eta^{(2)} = (\eta^{(1)})^{2}; \qquad \dots \; .
\end{equation*}
Observe that the equations for $\eta^{(1)}$ and $\eta^{(2)}$ are
linear equations, homogeneous for $\eta^{(1)}$ and nonhomogeneous
(with known right-hand side after $\eta^{(1)}$ has been found) for
$\eta^{(2)}$. Suppose that the first-order solution we are
interested in is $\eta^{(1)} = a e^{-i\omega_{0}t}$, already
introducing the complex arithmetic that will be used in the sequel
also. This solution is found for the initial values $\eta^{(1)}(0)
= a$, $\partial_{t} \eta^{(1)} (0) = -i \omega _{0} a$. Then the
equation for $\eta^{(2)}$, i.e. $\textmd{$\cal{L}$} \eta^{(2)} =
a^{2} e^{-2i\omega_{0}t}$ has as particular solution:
$\eta_{\textmd{p}}^{(2)} = A e^{-2i\omega_{0}t}$ with $A = -a^{2}
/(3 \omega_{0}^{2})$. This particular solution is the equivalent
of a `bound wave' mentioned above: it comes inevitably with the
first-order solution $\eta^{(1)}$. However,
$\eta_{\textmd{p}}^{(2)}$ will change the initial condition;
forcing it to remain unchanged could be done by adding a solution
$\eta_{\textmd{h}}^{(2)}$ of the homogeneous equation:
$\textmd{$\cal{L}$} \eta_{\textmd{h}}^{(2)} = 0$ that cancels the
particular solution at $t = 0$, explicitly:
$\eta_{\textmd{h}}^{(2)} = -\left( \frac{3}{2} A e^{-i\omega_{0}t}
- \frac{1}{2} A e^{i\omega_{0}t}\right)$. This homogeneous
solution corresponds to the second-order free wave mentioned
above. To avoid this solution to be present, the initial value has
to be taken like:
\begin{equation*}
\eta(0) = \epsilon a + \epsilon^{2}A; \qquad \partial_{t} \eta(0)
= -i \epsilon \omega_{0}a - 2i\epsilon^{2} \omega_{0} A.
\end{equation*}
The second-order terms in $\epsilon$ in these initial conditions
are similar to the second-order steering of the flap motion for the
signalling problem.\footnote[1]{Just as in this example, the
hierarchy of equations also continues for the \textsc{bvp}: there
will also be the third and higher order contributions, and bound and
free waves in each order. A higher-order steering than the second-order one
has not been done until now, since the effects are smaller,
although there are some exceptions. \par}

Besides the two difficult aspects of nature, dispersion and
nonlinearity, the precise description of the signal is also quite
involved, since the signal has to be described on a moving
boundary, the flap, which complicates matters also. For the rest of
this paper, we will describe the major details of this procedure.
The next section presents the \textsc{bvp} for the wave generation
problem. Section 3 and Section 4 discuss the first- and the
second-order wave generation theory, respectively. The final
section gives some conclusions about this paper. The results can
also be found in Dean and Dalrymple \cite{Dean} for the
first-order theory as well as in Sch\"{a}ffer \cite{Schaffer} for
the second-order theory, but our presentation is less technical
and emphasises the major steps.

\vspace{1.5cc}

\begin{center}
{\bf 2. GOVERNING EQUATION}
\end{center}

Let $\mathbf{u} = (u,w) = (\partial_{x} \phi, \partial_{z} \phi)$
define the velocity potential function $\phi = \phi(x,z,t)$ in a
Cartesian coordinate system $(x,z)$. Let also $\eta = \eta(x,t)$,
$\Xi = \Xi(z,t) = f(z)S(t)$, $g$, $h$, and $t$ denote surface wave
elevation, wavemaker position, gravitational acceleration, still
water depth and time, respectively. The governing equation for the
velocity potential is the Laplace equation
\begin{equation*}
  \partial_{x}^{2} \phi + \partial_{z}^{2} \phi = 0,
  \qquad \textmd{for}\; x \geq \Xi(z,t), \quad -h \leq z \leq \eta(x,t);
\end{equation*}
that results from the assumption that water (in a good
approximation) is incompressible: $\nabla \cdot \mathbf{u} = 0$.
The dynamic and kinematic free surface boundary conditions
(\textsc{dfsbc} and \textsc{kfsbc}), the kinematic boundary
condition at the wavemaker (\textsc{kwmbc}), and the bottom
boundary condition (\textsc{bbc}) are given by
\begin{equation*}
  \begin{array}{rrl}
    \textsc{dfsbc}: & \partial_{t} \phi + \frac{1}{2}|\nabla \phi|^{2} + g \eta = 0 & \qquad  \textrm{at} \; z = \eta(x,t); \\
    \textsc{kfsbc}: & \partial_{t} \eta + \partial_{x} \eta \partial_{x} \phi - \partial_{z} \phi = 0 & \qquad \textrm{at} \; z = \eta(x,t); \\
    \textsc{kwmbc}: & \partial_{x} \phi - f(z)S'(t) - f'(z)S(t) \partial_{z} \phi = 0 & \qquad \textrm{at} \; x = \Xi(z,t); \\
    \textsc{bbc}:   & \partial_{z} \phi = 0 & \qquad \textrm{at} \; z = -h. \\
  \end{array}
\end{equation*}
The \textsc{dfsbc} is obtained from Bernoulli's equation, the
\textsc{kfsbc} and the \textsc{kwmbc} are derived by applying the
material derivative to the surface elevation and wavemaker motion,
respectively. The \textsc{bbc} is obtained from the fact that water
neither comes in nor goes out of the wave tank. Note that the
\textsc{dfsbc} and \textsc{kfsbc} are nonlinear boundary
conditions prescribed at a yet unknown and moving free surface $z
= \eta(x,t)$. The elevation, potential and wavemaker position are
given by the following series expansions
\begin{eqnarray*}
  \eta &=& \epsilon \eta^{(1)} + \epsilon^{2} \eta^{(2)} + \epsilon^{3} \eta^{(3)} + \dots \\
  \phi &=& \epsilon \phi^{(1)} + \epsilon^{2} \phi^{(2)} + \epsilon^{3} \phi^{(3)} + \dots \\
     S &=& \epsilon    S^{(1)} + \epsilon^{2}    S^{(2)} + \epsilon^{3}    S^{(3)} + \dots,
\end{eqnarray*}
where $\epsilon$ is a small parameter, a measure of the surface
elevation nonlinearity.
\begin{figure}[h]
  \begin{center}
    \epsfig{file=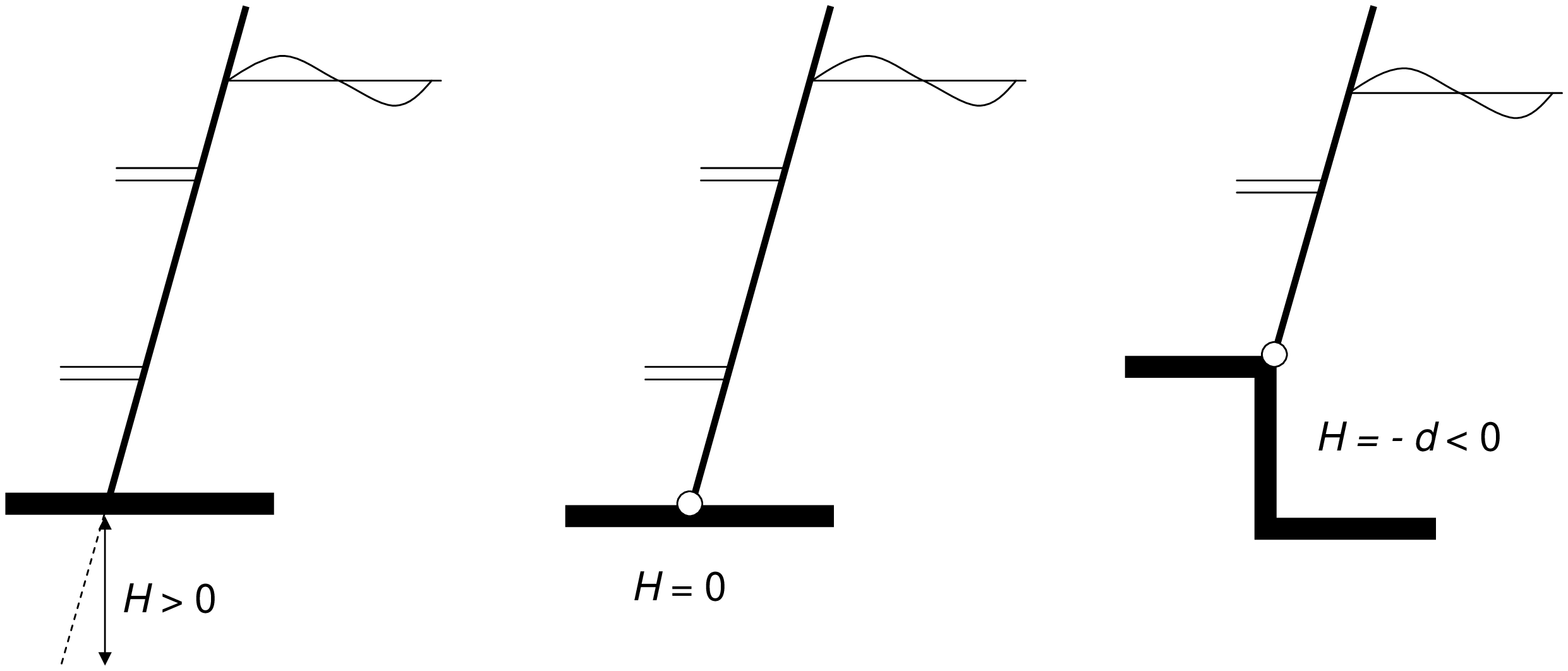, angle = -90, width = 0.9\textwidth}
  \end{center}
  \caption{\footnotesize The flap type of wavemaker with different center
  of rotations: below the bottom (left), at the bottom (middle), and
  above the bottom (right).} \label{flapgeometry}
\end{figure}

The wavemaker we will consider is a rotating flap, see Figure
\ref{flapgeometry}. It is given by $\Xi(z,t) = f(z)S(t)$, where
$f(z)$ describes the geometry of the wavemaker:
\begin{equation}
  f(z) = \left\{
\begin{array}{ll}
{\displaystyle
    1 + \frac{z}{h + H}}, & \qquad \hbox{for \; $-(h - d) \leq z \leq 0$;} \\
    0,                    & \qquad \hbox{for \; $-h \leq z < -(h - d)$.}   \\
\end{array}
\right. \label{wavemakertype}
\end{equation}
Note that $f(z)$ is given by design, and $S(t)$ is the wavemaker
motion that can be controlled externally to generate different
types of waves. The center of rotation is at $z = -(h + H)$. If
the center of rotation is at or below the bottom, then $d = 0$ and
in fact, we do not have the second case of (\ref{wavemakertype}). If
the center of rotation is at a height $d$ above the bottom, then
$d = -H$.
\vspace{1.5cc}
\begin{center}
{\bf 3. FIRST-ORDER WAVE GENERATION THEORY}
\end{center}

In this section, we solve a homogeneous \textsc{bvp} for the
first-order wave generation theory. By prescribing the first-order
wavemaker motion as a linear superposition of monochromatic
frequencies, we find the generated surface elevation also as a
linear superposition of monochromatic modes. After applying the
Taylor series expansion of the potential function $\phi$ around $x
= 0$ and $z = 0$ as well as applying the series expansion method,
the first-order potential function has to satisfy the Laplace
equation
\begin{equation}
  \partial_{x}^{2} \phi^{(1)} + \partial_{z}^{2} \phi^{(1)} = 0,
  \qquad \textmd{for}\; x \geq 0, \;\; -h \leq z \leq 0.
  \label{1stLaplace}
\end{equation}
We also obtain the \textsc{bvp} for the first-order wave
generation theory at the lowest expansion order. It reads
\begin{equation}
  \begin{array}{rcll}
    g \eta^{(1)} + \partial_{t} \phi^{(1)} &=& 0,             \qquad & \textmd{at} \; z = 0; \\
    \partial_{t} \eta^{(1)} - \partial_{z} \phi^{(1)} &=& 0,  \qquad & \textmd{at} \; z = 0; \\
    \partial_{x} \phi^{(1)} - f(z) {\displaystyle \frac{dS^{(1)}}{dt}} &=& 0, \qquad & \textmd{at} \; x = 0; \\
    \partial_{z} \phi^{(1)} & = & 0,                          \qquad & \textmd{at} \; z = -h. \label{firstorderBCs}
  \end{array}
\end{equation}
By combining the \textsc{dfsbc} and the \textsc{dfsbc} at $z = 0$
(\ref{firstorderBCs}), we obtain the first-order homogeneous free
surface boundary condition
\begin{equation}
  g \partial_{z} \phi^{(1)} + \partial_{t}^{2} \phi^{(1)} = 0,
  \qquad \textrm{at} \; z = 0.  \label{combinedfsc}
\end{equation}

We look for the so-called monochromatic waves
\begin{equation*}
  \phi^{(1)}(x,z,t) = \psi(z) e^{-i \theta(x,t)},
\end{equation*}
where $\theta(x,t) = k x - \omega t$. Then from the Laplace
equation (\ref{1stLaplace}), we have $\psi''(z) - k \psi(z) = 0$,
for $-h \leq z \leq 0$. Applying the \textsc{bbc} leads to
$\psi(z) = \alpha \cosh k(z + h)$, $\alpha \in \mathbb{C}$. From
the combined free surface condition (\ref{combinedfsc}), we obtain
a relation between the wavenumber $k$ and frequency $\omega$,
known as the \textsl{linear dispersion relation} (\textsc{ldr}),
explicitly given by
\begin{equation}
  \omega^{2} = g k \tanh k h.  \label{ldr}
\end{equation}

Let us assume that the first-order wavemaker motion $S^{(1)}(t)$
is given by a harmonic function with frequency $\omega_{n}$ and
maximum stroke $|S_{n}|$ from an equilibrium position, represented
in complex notation as
\begin{equation*}
  S^{(1)}(t) = \sum_{n = 1}^{\infty} -\frac{1}{2}i S_{n} e^{i \omega_{n}t} + \textmd{c.c.},
\end{equation*}
where c.c. denotes the complex conjugate of the preceding term.
Since this `first-order steering' contains an infinite number of
discrete frequencies $\omega_{n}$, it motivates us to write a
general solution for the potential function by linear
superposition of discrete spectrum. By choosing the arbitrary
spectral coefficient ${\displaystyle \alpha  = \frac{ig}{2\omega_{n}}
\frac{C_{n}}{\cosh k_{n}h}}$, the first potential function is found
to be
\begin{equation*}
  \phi^{(1)}(x,z,t) = \sum_{n = 1}^{\infty} \frac{ig}{2\omega_{n}} C_{n}
  \frac{\cosh k_{n}(z + h)}{\cosh k_{n} h} e^{-i\theta_{n}(x,t)} + \textmd{c.c.},
\end{equation*}
where $\theta_{n}(x,t) = k_{n} x - \omega_{n}t$, with
wavenumber-frequency pairs $(k_{n},\omega_{n})$, $n \in
\mathbb{Z}$ satisfying the \textsc{ldr} (\ref{ldr}). For a
continuous spectrum, the summation is replaced by an integral.
Allowing the wavenumber to be complex valued, the \textsc{ldr}
becomes
\begin{equation*}
  \omega_{n}^{2} = g k_{nj} \tanh k_{nj}h, \qquad j \in \mathbb{N}_{0}.
\end{equation*}
For $j = 0$, the wavenumber is real and it corresponds to the
propagating mode of the surface wave elevation. For $j \in
\mathbb{N}$, the wavenumbers are purely imaginary, and thus $i
k_{nj} \in \mathbb{R}$. Since we are interested in the decaying
solution, we choose $i k_{nj} > 0$ and hence the modes of these
wavenumbers are called the \textsl{evanescent modes}. As a consequence, the
first-order potential function can now be written as
\begin{equation}
  \phi^{(1)}(x,z,t) = \sum_{n = 1}^{\infty} \sum_{j =  0}^{\infty} \frac{i g}{2 \omega_{n}} C_{nj}
  \frac{\cosh k_{nj}(z + h)}{\cosh k_{nj}h} e^{-i \theta_{nj}(x,t)} + \textmd{c.c.}, \label{1stpotential}
\end{equation}
where $\theta_{nj}(x,t) = k_{nj}x - \omega_{n}t$.
\begin{figure}[h]
  \begin{center}
    \epsfig{file=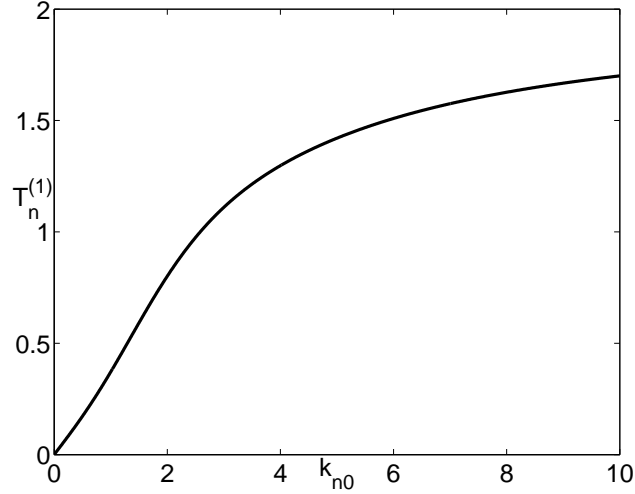, width = 0.5\textwidth}
  \end{center}
  \caption{\footnotesize The first-order transfer function plot as a function
  of wavenumber $k_{n0}$, for the case that the water depth is $h = 1$ and the
  center of rotation is at $d = \frac{1}{3} h$ above the tank floor.} \label{gambarTF1}
\end{figure}

Furthermore, applying the \textsc{kwmbc} (\ref{firstorderBCs}),
integrating along the water depth, and using the property that
$\Big\{ \cosh k_{nj}(z + h), \cosh k_{nl}(z + h), \; j, l \in
\mathbb{N}_{0} \Big\}$ is a set of orthogonal functions for $j
\neq l$, we can find the surface wave complex-valued amplitude
$C_{nj}$ as follows
\begin{eqnarray*}
  C_{nj} &=& \frac{\omega_{n}^{2} S_{n}}{g k_{nj}} \cosh k_{nj}h
      \frac{\displaystyle \int_{-h}^{0} f(z) \cosh k_{nj}(z + h)\,dz}{\displaystyle \int_{-h}^{0} \cosh^{2} k_{nj}(z + h)\,dz} \nonumber \\
  &=& \frac{4 S_{n} \sinh k_{nj}h}{k_{nj}(h + H)}\,
      \frac{k_{nj}(h + H) \sinh k_{nj}h + \cosh k_{nj}d - \cosh k_{nj}h}{2k_{nj}h + \sinh (2k_{nj}h)}, \qquad j \in \mathbb{N}_{0}.
\end{eqnarray*}
Finally, the first-order surface elevation can be found from the
\textsc{dfsbc} (\ref{firstorderBCs}), and is given as follows:
\begin{equation*}
  \eta^{(1)}(x,t) = \sum_{n = -\infty}^{\infty} \sum_{j =  0}^{\infty}
  \frac{1}{2} C_{nj} e^{-i \theta_{nj}(x,t)} + \textmd{c.c.}
\end{equation*}
This first-order theory can also be found in Dean and Dalrymple
\cite{Dean}.\newpage
\begin{center}
  \textsc{Remark 1.}
\end{center}
For `practical' purposes, it is useful to introduce the so-called
\textit{transfer function} or \textit{frequency response} of a
system. It is defined as the ratio of the output and the input of
a system. In our wave generation problem, we have a system with a
wavemaker motion as input and the surface wave amplitude as
output. Therefore, the first-order transfer function $T_{n}^{(1)}$
is defined as the ratio between the surface wave amplitude of the
propagating mode $C_{n0}$ as output and the maximum stroke
$|S_{n}|$ as input, explicitly given by
\begin{equation*}
  T_{n}^{(1)} = 4 \; \frac{\sinh k_{n0}h}{k_{n0}(h + H)} \;
  \frac{k_{n0}(h + H) \sinh k_{n0}h + \cosh k_{n0}d - \cosh k_{n0}h}{2 k_{n0}h + \sinh 2k_{n0}h}.
\end{equation*}
Figure \ref{gambarTF1} shows the first-order transfer function
plot as function of wavenumber $k_{n0}$ for a given water depth
$h$ and the center of rotation $d$. For increasing $k_{n0}$, which
also means increasing frequency $\omega_{n}$, the transfer
function is monotonically increasing as well. It increases faster
for smaller values of $k_{n0}$ and slower for larger values of
$k_{n0}$, approaching the asymptotic limit of $T_{n}^{(1)} = 2$
for $k_{n0}h \rightarrow \infty$.

\vspace{1.5cc}

\begin{center}
{\bf 4. SECOND-ORDER WAVE GENERATION THEORY}
\end{center}

In this section, we solve a nonhomogeneous \textsc{bvp} for the
second-order wave generation theory. Due to the nonhomogeneous
boundary condition at the free surface, which causes interactions
between each possible pair of first-order wave components, the
resulting surface wave elevation has a second-order effect, known
as the bound wave component. Furthermore, due to the first-order
wavemaker motion and the boundary condition at the wavemaker, the
generated wave also has another second-order effect, namely the
free wave component. The latter component is undesired since it
results in a spatially inhomogeneous wave field due to the
different propagation velocities of bound wave and free wave
components with the same frequency. Therefore, in order to prevent
the free wave component to be generated, we add an additional
second-order bound wave effect to the flap motion. This process is
known as `second-order steering' of the wavemaker motion. More
details about this theory, including an experimental verification,
can be found in Sch\"{a}ffer \cite{Schaffer}. For the history of
wave generation theory, see also references in this paper.

Taking terms of the second-order in the series expansion, we
obtain the \textsc{bvp} for the second-order wave generation
theory. The second-order potential function also satisfies the
Laplace equation
\begin{equation*}
  \partial_{x}^{2} \phi^{(2)} + \partial_{z}^{2} \phi^{(2)} = 0,
  \qquad \textmd{for}\; x \geq 0, \;\; -h \leq z \leq 0.
\end{equation*}
Almost all the second-order boundary conditions now become
nonhomogeneous
\begin{equation}
  \begin{array}{rcll}
  g \eta^{(2)} + \partial_{t} \phi^{(2)}
  &=& - \Big(\eta^{(1)} \partial_{tz}^{2} \phi^{(1)} + \frac{1}{2} |\nabla \phi^{(1)}|^{2} \Big), & \qquad \textmd{at} \; z = 0; \\
  \partial_{t} \eta^{(2)} - \partial_{z} \phi^{(2)}
  &=& \eta^{(1)} \partial_{z}^{2} \phi^{(1)} - \partial_{x} \eta^{(1)} \partial_{x} \phi^{(1)},   & \qquad \textmd{at} \; z = 0; \\
  \partial_{x} \phi^{(2)} - f(z) {\displaystyle \frac{dS^{(2)}}{dt}}
  &=& S^{(1)}(t) \Big(f'(z) \partial_{z} \phi^{(1)} - f(z) \partial_{x}^{2} \phi^{(1)} \Big),     & \qquad \textmd{at} \; x = 0; \\
    \partial_{z} \phi^{(2)} &=& 0, & \qquad \textmd{at} \; z = -h. \label{secondorderBCs}
  \end{array}
\end{equation}
By combining the \textsc{dfsbc} and \textsc{kfsbc} of
(\ref{secondorderBCs}) at $z = 0$, we have the second-order
nonhomogeneous free surface boundary condition
\begin{equation}
  g \partial_{z} \phi^{(2)} + \partial_{t}^{2} \phi^{(2)} = \textsc{rhs}_{1}, \qquad \textrm{at} \; z = 0.  \label{FSDKBC}
\end{equation}
Using the first-order potential function (\ref{1stpotential}),
\textsc{rhs}$_{1}$ is explicitly given by
\begin{eqnarray*}
  \textsc{rhs}_{1} &=& \left. -\left(\frac{\partial}{\partial t} |\nabla \phi^{(1)}|^{2} + \eta^{(1)} \frac{\partial}{\partial z}
  \left[g \frac{\partial \phi^{(1)}}{\partial z} + \frac{\partial^{2} \phi^{(1)}}{\partial t^{2}} \right] \right) \right|_{z = 0} \nonumber \\
  &=& \sum_{m,n = 1}^{\infty} \sum_{l,j =  0}^{\infty}
  \left(A_{mnlj}^{+} e^{-i(\theta_{ml} + \theta_{nj})} + A_{mnlj}^{-} e^{-i(\theta_{ml} - \theta_{nj}^{\ast})} \right) + \textmd{c.c.},
\end{eqnarray*}
where
\begin{eqnarray*}
  \frac{A_{mnlj}^{+}}{C_{ml} C_{nj}} \!\!\! &=& \!\!\!
                     \frac{1}{4i} \!\! \left[(\omega_{m} + \omega_{n}) \! \left(g^{2} \frac{k_{ml} k_{nj}}{\omega_{m} \omega_{n}}
                   - \omega_{m} \omega_{n} \right)
                   + \frac{g^{2}}{2} \! \left(\frac{k_{ml}^{2}}{\omega_{m}} + \frac{k_{nj}^{2}}{\omega_{n}} \right)
                   - \frac{1}{2} (\omega_{m}^{3} + \omega_{n}^{3}) \right]\!\!, \\
  \frac{A_{mnlj}^{-}}{C_{ml} C_{nj}^{\ast}} \!\!\! &=& \!\!\!
                     \frac{1}{4i} \!\! \left[(\omega_{m} - \omega_{n}) \! \left(g^{2} \frac{k_{ml} k_{nj}^{\ast}}{\omega_{m} \omega_{n}}
                   + \omega_{m} \omega_{n} \right)
                   + \frac{g^{2}}{2} \! \left(\frac{k_{ml}^{2}}{\omega_{m}} - \frac{k_{nj}^{2}}{\omega_{n}} \right)
                   - \frac{1}{2} (\omega_{m}^{3} - \omega_{n}^{3}) \right]\!\!.
\end{eqnarray*}

In order to find the bound wave component, the free wave component,
and to apply the second-order steering wavemaker motion, we split
the second-order \textsc{bvp} (\ref{secondorderBCs}) into three
\textsc{bvp}s. For that purpose, the second-order potential
function is split into three components as follows
\begin{equation*}
  \phi^{(2)}(x,z,t) = \phi^{(21)}(x,z,t) + \phi^{(22)}(x,z,t) + \phi^{(23)}(x,z,t).
\end{equation*}
Now the corresponding \textsc{bvp} for the first component of the
potential function $\phi^{(21)}$ reads
\begin{equation}
  \begin{array}{rcll}
    g \partial_{z} \phi^{(21)} + \partial_{t}^{2} \phi^{(21)} &=& \textsc{rhs}_{1},  & \qquad \textrm{at} \; z = 0;  \\
      \partial_{z} \phi^{(21)} &=& 0,                                                & \qquad \textrm{at} \; z = -h. \label{1stcomponent_bvp}
  \end{array}
\end{equation}
The corresponding \textsc{bvp} for the second component of the
potential function $\phi^{(22)}$ reads
\begin{equation}
  \begin{array}{rcll}
    g \partial_{z} \phi^{(22)} + \partial_{t}^{2} \phi^{(22)} &=& 0, & \qquad \textrm{at} \; z = 0; \\
    \partial_{x} \phi^{(22)} &=& S^{(1)}(t) \Big(f'(z) \partial_{z} \phi^{(1)} - f(z) \partial_{x}^{2} \phi^{(1)} \Big)
    - \partial_{x} \phi^{(21)},     & \qquad \textmd{at} \; x = 0;                   \label{2ndcomponent_bvp} \\
    \partial_{z} \phi^{(22)} &=& 0, & \qquad \textrm{at} \; z = -h.
  \end{array}
\end{equation}
And the \textsc{bvp} for the third component of the potential
function $\phi^{(23)}$ reads
\begin{equation}
  \begin{array}{rcll}
    g \partial_{z} \phi^{(23)} + \partial_{t}^{2} \phi^{(23)} &=& 0,         & \qquad \textrm{at} \; z = 0; \\
    \partial_{x} \phi^{(23)} &=& f(z) {\displaystyle \frac{dS^{(2)}}{dt}},   & \qquad \textrm{at} \; x = 0; \\
    \partial_{z} \phi^{(23)} &=& 0,                                          & \qquad \textrm{at} \; z = -h. \label{3rdcomponent_bvp}
  \end{array}
\end{equation}

By taking the Ansatz for the first part of the second-order
potential function $\phi^{(21)}$ as follows
\begin{eqnarray*}
  \phi^{(21)}(x,z,t) &=& \sum_{m,n = 1}^{\infty} \sum_{l,j =  0}^{\infty}
      B_{mnlj}^{+} \frac{\cosh (k_{ml} + k_{nj})(z + h)}       {\cosh (k_{ml} + k_{nj})h}        e^{-i(\theta_{ml} + \theta_{nj})} \nonumber \\
  &+& B_{mnlj}^{-} \frac{\cosh (k_{ml} - k_{nj}^{\ast})(z + h)}{\cosh (k_{ml} - k_{nj}^{\ast})h} e^{-i(\theta_{ml} - \theta_{nj}^{\ast})}
  + \textmd{c.c.},
\end{eqnarray*}
then we can derive the corresponding coefficients to be:
\begin{eqnarray*}
  B_{mnlj}^{+} &=& \frac{A_{mnlj}^{+}}{\Omega^{2}(k_{ml} + k_{nj})        - (\omega_{m} + \omega_{n})^{2}}, \\
  B_{mnlj}^{-} &=& \frac{A_{mnlj}^{-}}{\Omega^{2}(k_{ml} - k_{nj}^{\ast}) - (\omega_{m} - \omega_{n})^{2}}.
\end{eqnarray*}

This first component of the second-order potential function will
contribute the bound wave component to the second-order surface
wave elevation $\eta^{(2)}$. For $j = 0$, the wave component is a
propagating mode and for $j \in \mathbb{N}$, it consists of
evanescent modes. Since the wavenumbers $k_{mj} + k_{nj}$ and
$k_{mj} + k_{nj}^{\ast}$, $j \in \mathbb{N}_{0}$ do not satisfy
the \textsc{ldr} with frequencies $\omega_{m} \pm \omega_{n}$,
then the denominator part of $B_{mnj}^{\pm}$ will never vanish and
thus the potential function is a bounded function.

Let the right-hand side of the boundary condition at the wavemaker
for the second \textsc{bvp} (\ref{2ndcomponent_bvp}) be denoted by
\textsc{rhs}$_{2}$, which is expressed as
\begin{equation*}
  \textsc{rhs}_{2} = \sum_{m,n = 1}^{\infty} \sum_{l,j =  0}^{\infty}
      \left(F_{mnlj}^{+}(z) e^{i(\omega_{m} + \omega_{n})t} + F_{mnlj}^{-}(z) e^{i(\omega_{m} - \omega_{n})t} \right) + \textmd{c.c.},
\end{equation*}
where
\begin{eqnarray*}
  F_{mnlj}^{+}(z) &=& \frac{g}{8 \omega_{n}} \frac{S_{m} k_{nj} C_{nj}}{\cosh k_{nj}h}
  \left[f'(z) \sinh k_{nj}(z + h) + k_{nj} f(z) \cosh k_{nj}(z + h) \right] \nonumber \\
  && + \; i (k_{ml} + k_{nj}) B_{mnlj}^{+} \frac{\cosh (k_{ml} + k_{nj})(z + h)}{\cosh (k_{ml} + k_{nj})h}, \\
  F_{mnlj}^{-}(z) &=& -\frac{g}{8 \omega_{n}} \frac{S_{m} k_{nj}^{\ast} C_{nj}^{\ast}}{\cosh k_{nj}^{\ast}h}
  \left[f'(z) \sinh k_{nj}^{\ast}(z + h) + k_{nj}^{\ast} f(z) \cosh k_{nj}^{\ast}(z + h) \right] \nonumber \\
  && + \; i (k_{ml} - k_{nj}^{\ast}) B_{mnlj}^{-} \frac{\cosh (k_{ml} - k_{nj}^{\ast})(z + h)}{\cosh (k_{ml} - k_{nj}^{\ast})h}.
\end{eqnarray*}
Let the Ansatz for the second component of the second-order
potential function $\phi^{(22)}$ be
\begin{eqnarray*}
  \phi^{(22)}(x,z,t) \!\! &=& \!\!\!\!\! \sum_{m,n = 1}^{\infty} \sum_{l,j =  0}^{\infty}
  \left(\frac{i\,g\,P_{mnlj}^{+} }{2(\omega_{m} + \omega_{n})}  \frac{\cosh K_{mnlj}^{+}(z + h)}{\cosh K_{mnlj}^{+}h}
  e^{-i(K_{mnj}^{+}x - (\omega_{m} + \omega_{n})t)} \right. \nonumber \\
  &+& \left. \frac{i\,g\,P_{mnlj}^{-}}{2(\omega_{m} - \omega_{n})} \frac{\cosh K_{mnlj}^{-}(z + h)}{\cosh K_{mnlj}^{-}h}
  e^{-i(K_{mnlj}^{-}x - (\omega_{m} - \omega_{n})t)} \right) + \textmd{c.c.},
\end{eqnarray*}
where the wavenumbers $K_{mnlj}^{\pm}$, $j \in \mathbb{N}_{0}$ and
frequencies $\omega_{m} \pm \omega_{n}$ satisfy the \textsc{ldr}.
Using the property that $\Big\{\cosh K_{mnlj}^{\pm}(z + h), \cosh
K_{mnl'j'}^{\pm}(z + h), \; l, l', j, j' \in \mathbb{N}_{0}
\Big\}$ is a set of orthogonal functions for $l \neq l'$ and $j
\neq j'$, we find the coefficients $P_{mnlj}^{\pm}$ as follow
\begin{eqnarray}
  P_{mnlj}^{\pm} &=& \frac{2(\omega_{m} \pm \omega_{n}) \cosh K_{mnlj}^{\pm}h}{g K_{mnlj}^{\pm}}
  \frac{\displaystyle \int_{-h}^{0} F_{mnlj}^{\pm}(z) \cosh K_{mnlj}^{\pm} (z + h)\,dz}
  {\displaystyle \int_{-h}^{0} \cosh^{2} K_{mnlj}^{\pm} (z + h)\,dz} \nonumber \\
  &=& 8 \frac{K_{mnlj}^{\pm} \sinh K_{mnlj}^{\pm}h}{\omega_{m} \pm \omega_{n}}
  \frac{\displaystyle \int_{-h}^{0} F_{mnlj}^{\pm}(z) \cosh K_{mnlj}^{\pm} (z + h)\,dz}{2 K_{mnlj}^{\pm} h + \sinh (2 K_{mnlj}^{\pm} h)}.
  \label{2ndcomponent_potential}
\end{eqnarray}
The second component of the second-order potential function
$\phi^{(22)}$ will give contributions to the free wave component
of the second-order surface wave elevation $\eta^{(2)}$. This
component arises due to the boundary condition at the wavemaker
caused by the first-order wavemaker motion. Since the desired
surface elevation is only the bound wave component, we want to get
rid this term, especially the propagating mode. The evanescent
modes vanish anyway after they evolve far away from the
wavemaker. By prescribing the second-order wavemaker motion such
that the propagating mode of the third component $\phi^{(23)}$
will cancel the same mode of the second one $\phi^{(22)}$, then
far from the wavemaker we have the desired bound wave component
only.

Let the second-order wavemaker motion be given by
\begin{equation*}
  S^{(2)}(t) = \sum_{m,n = 1}^{\infty} -\frac{1}{2} i \left(S_{mn}^{+} e^{i (\omega_{m} + \omega_{n})t}
  + S_{mn}^{-} e^{i (\omega_{m} - \omega_{n})t} \right) + \textmd{c.c.}
\end{equation*}
Let also the Ansatz for the third component of the second-order
potential function $\phi^{(23)}$ be
\begin{eqnarray*}
  \phi^{(23)}(x,z,t) \!\! &=& \!\!\!\!\! \sum_{m,n = 1}^{\infty} \sum_{l,j =  0}^{\infty}
  \left(\frac{i\,g\,Q_{mnlj}^{+}}{2(\omega_{m} + \omega_{n})} \frac{\cosh K_{mnlj}^{+}(z + h)}{\cosh K_{mnlj}^{+}h}
  e^{-i(K_{mnlj}^{+}x - (\omega_{m} + \omega_{n})t)} \right. \nonumber \\
  &+& \left. \frac{i\,g\,Q_{mnlj}^{-}}{2(\omega_{m} - \omega_{n})} \frac{\cosh K_{mnlj}^{-}(z + h)}{\cosh K_{mnlj}^{-}h}
  e^{-i(K_{mnlj}^{-}x - (\omega_{m} - \omega_{n})t)} \right) + \textmd{c.c.},
\end{eqnarray*}
where $\Omega(K_{mnlj}^{\pm}) = \omega_{m} \pm \omega_{n}$. Using
the orthogonality property again, we find the coefficients
$Q_{mnlj}^{\pm}$ as follows
\begin{eqnarray}
  Q_{mnlj}^{\pm} &=& S_{mn}^{\pm} \sinh K_{mnlj}^{\pm}h
  \frac{\displaystyle \int_{-h}^{0} f(z) \cosh K_{mnlj}^{\pm} (z + h)\,dz}
  {\displaystyle \int_{-h}^{0} \cosh^{2} K_{mnlj}^{\pm} (z + h)\,dz} \nonumber \\
  &=& \frac{4 S_{mn}^{\pm} \sinh K_{mnlj}^{\pm}h}{K_{mnlj}^{\pm}(h + H)} \,
  \frac{K_{mnlj}^{\pm}(h + H) \sinh K_{mnlj}^{\pm}h + \cosh K_{mnlj}^{\pm}d - \cosh K_{mnlj}^{\pm}h}
  {2 K_{mnlj}^{\pm} h + \sinh (2 K_{mnlj}^{\pm} h)}. \qquad \qquad \label{3rdcomponent_potential}
\end{eqnarray}
To have the propagating mode of the free wave from the second
$\big(\phi^{(22)} \big)$ and the third $\big(\phi^{(23)} \big)$
components cancel each other, we must require $P_{mn00} + Q_{mn00}
= 0$, which leads to the following second-order wavemaker motion,
known as `second-order steering':
\begin{equation*}
  S_{mn}^{\pm} = \frac{2 \big(K_{mn00}^{\pm} \big)^{2} I_{mn00}^{\pm} (h + H)}
  {\big(\omega_{m} \pm \omega_{n} \big)\big(\cosh K_{mn00}^{\pm}h - \cosh K_{mn00}^{\pm}d - K_{mn00}^{\pm}(h + l) \sinh K_{mn00}^{\pm}h\big)},
\end{equation*}
where
\begin{equation*}
  I_{mn00}^{\pm} = \int_{-h}^{0} F_{mn00}^{\pm}(z) \cosh K_{mn00}^{\pm}(z + h)\,dz.
\end{equation*}
Therefore, with this choice of the second-order wavemaker motion, the
second-order potential function can be written as
\begin{equation*}
  \phi^{(2)}(x,z,t) = \phi^{(2)}_{\textmd{\tiny propagating}}(x,z,t) + \phi^{(2)}_{\textmd{\tiny evanescent}}(x,z,t),
\end{equation*}
where
\begin{eqnarray*}
  \phi^{(2)}_{\textmd{\tiny propagating}} &=& \phi^{(21)}_{\textmd{\tiny bound wave, propagating}}, \\
  \phi^{(2)}_{\textmd{\tiny evanescent}}  &=& (\phi^{(21)}_{\textmd{\tiny bound wave}} +
  \phi^{(22)}_{\textmd{\tiny free wave}}   +   \phi^{(23)}_{\textmd{\tiny free wave}})_{\textmd{\tiny evanescent}}.
\end{eqnarray*}
Consequently, from the second-order \textsc{dfsbc}
(\ref{secondorderBCs}), we find the second-order surface wave
elevation. It can be written as follows
\begin{equation*}
    \eta^{(2)}(x,t) = \eta^{(2)}_{\textmd{\tiny propagating}} + \phi^{(2)}_{\textmd{\tiny evanescent}},
\end{equation*}
where
\begin{equation*}
  \eta^{(2)}_{\textmd{\tiny propagating}} = \sum_{m,n = 1}^{\infty}
  D_{mn00}^{+} e^{-i(\theta_{m0} + \theta_{n0})} + D_{mn00}^{-} e^{-i(\theta_{m0} - \theta_{n0})},
\end{equation*}
and
\begin{eqnarray*}
  \eta^{(2)}_{\textmd{\tiny evanescent}} &=& \sum_{m,n = 1}^{\infty} \sum_{j = 1, \atop l = 0}^{\infty}
  \left(D_{mnlj}^{+} e^{-i(\theta_{ml} + \theta_{nj})} +
        D_{mnlj}^{-} e^{-i(\theta_{ml} - \theta_{nj}^{\ast})} \right. \nonumber \\
  &+&        \frac{1}{2} (P_{mnlj}^{+} + Q_{mnlj}^{+}) e^{-i(K_{mnlj}^{+}x - (\omega_{m} + \omega_{n})t)} \nonumber \\
  &+& \left. \frac{1}{2} (P_{mnlj}^{-} + Q_{mnlj}^{-}) e^{-i(K_{mnlj}^{-}x - (\omega_{m} - \omega_{n})t)} \right)
   +  \textmd{c.c.},
\end{eqnarray*}
where for $l,j \in \mathbb{N}_{0}$:
\begin{eqnarray*}
  D_{mnlj}^{+} &=& -\frac{1}{g} \left[i(\omega_{m} + \omega_{n})B_{mnlj}^{+}
                +  \frac{1}{8} \left(g^{2} \frac{k_{ml} k_{nj}}{\omega_{m} \omega_{n}}
                -  \omega_{m} \omega_{n} - (\omega_{m}^{2} + \omega_{n}^{2}) \right) C_{ml} C_{nj} \right], \\
  D_{mnlj}^{-} &=& -\frac{1}{g} \left[i(\omega_{m} - \omega_{n})B_{mnlj}^{-}
                +  \frac{1}{8} \left(g^{2} \frac{k_{ml} k_{nj}^{\ast}}{\omega_{m} \omega_{n}}
                +  \omega_{m} \omega_{n} - (\omega_{m}^{2} + \omega_{n}^{2}) \right) C_{ml} C_{nj}^{\ast} \right].
\end{eqnarray*}

We have seen that the first-order surface wave elevation consists
of a linear superposition of monochromatic frequencies. However,
due to nonlinear effects, nonhomogeneous \textsc{bvp}, and
interactions of the first-order wave components, the second-order
surface elevation is composed by a superposition of bichromatic
frequencies $\omega_{m} \pm \omega_{n}$. The components with
frequency $\omega_{m} + \omega_{n}$ are called the
`\textsl{superharmonics}' and those with frequency $|\omega_{m} -
\omega_{n}|$ are called the `\textsl{subharmonics}'.

\begin{center}
  \textsc{Remark 2.}
\end{center}
Similar to the first-order wave generation theory, we can define a
second-order transfer function as well. A detailed formula for this
transfer function can be found in Sch\"{a}ffer~\cite{Schaffer}.

\vspace{1.5cc}

\begin{center}
{\bf 5. CONCLUSIONS}
\end{center}

We have discussed the theory for wave generation based on the
fully nonlinear water wave equation. We solved a nonlinear
\textsc{bvp} by the series expansion method. Using this method,
the problem turns into a set of linear \textsc{bvp}s at each
expansion order. The lowest order gives a homogeneous \textsc{bvp}
and the higher orders give nonhomogeneous ones with known,
depending on previous solutions, right-hand sides. 
In this paper, we focussed on the wave generation theory up to the second-order.

Based on the first-order wave generation theory, we describe the
surface wave fields as the superposition of monochromatic waves.
Due to the \textsc{bvp}, the wavenumbers and frequencies of this
wave field are related by the \textsc{ldr}. By prescribing the
first-order steering of the wavemaker as a linear superposition of
harmonic motions, we found that the first-order surface elevation
is simply a linear superposition of the corresponding
monochromatic waves. Furthermore, the wavemaker transfer function
is also introduced for practical purposes in the laboratory.

For the second-order wave generation theory, we have solved a
nonhomogeneous \textsc{bvp}. Due to the interactions between each
pair of first-order wave components, the second-order wave field
has bound wave components. Additionally, due to the first-order
steering of the wavemaker motion and the boundary condition at the
wavemaker, a free wave component is also generated, which is undesired. 
Therefore, we prevent it by controlling this second-order wavemaker motion. 
By applying this second-order steering, the resulting surface wave field
contains only the desired bound wave components. Similarly, one
can find the second-order transfer function for the relationship between
the first-order and the second-order motions.

\vspace{1.5cc} \noindent{\bf Acknowledgement. }{\small This
research has been executed partly in Indonesia and in the
Netherlands. In Indonesia, it was done during a visit at
Industrial and Applied Mathematics Research and Development Group,
Institut Teknologi Bandung (\textsc{kpp-mit itb}). In the
Netherlands, it was done at the University of Twente. We gratefully
acknowledge both the Small Project Facility of the European Union
Jakarta, entitled \textit{`Building Academia-Industry Partnership
in the Sectors of Marine and Telecommunication Technology'} and
the project \textit{`Prediction and Generation of Deterministic
Extreme Waves in Hydrodynamic Laboratories'}
(\textsc{twi}.\textsf{5374}) of the Netherlands Organization of
Scientific Research \textsc{nwo}, subdivision Applied Sciences
\textsc{stw}. We also appreciate the fruitful discussions with
Professor E. (Brenny) van Groesen and Gert Klopman at the University of Twente. 
The financial assistance for the publication of this paper from the 
Faculty of Engineering, The University of Nottingham, University Park and Malaysia Campuses 
under the New Researcher Fund NRF.5035 is also greatly acknowledged.\par}

\vspace{2cc}
\begin{center}
{\small\bf REFERENCES}
\end{center}

\newcounter{ref}
\begin{list}{\small \arabic{ref}.}{\usecounter{ref} \leftmargin 4mm
\itemsep -1mm}
\bibitem{Dean} {\small {\sc R. G. Dean} and {\sc R. A. Dalrymple},
{\it Water Wave Mechanics for Engineers and Scientists}, volume
{\bf 2} of {\it Advanced Series of Ocean Engineering}. World
Scientific, Singapore, 1991.}

\bibitem{Schaffer} {\small {\sc H. A. Sch\"{a}ffer}, ``Second-order wavemaker theory for irregular waves'',
{\it Ocean Engng.} {\bf 23} (1996), 1:47--88.}
\end{list}

\vspace{1cc}
{\small\noindent {\sd Natanael Karjanto}:
Department of Applied Mathematics, University of Twente, Postbus 217, 7500 AE, Enschede, The Netherlands.\\
\noindent E-mail: \url{n.karjanto@math.utwente.nl}
\vspace{0.75cc} \\
Department of Applied Mathematics, Faculty of Engineering, The University of Nottingham Malaysia Campus,
Jalan Broga, Semenyih 43500, Selangor, Malaysia.\\
\noindent E-mail: \url{natanael.karjanto@nottingham.edu.my}
}
\end{document}